\documentstyle[prd,aps,preprint]{revtex}

\tighten
\def\bort#1{}

\def\be{\begin{equation}}
\def\bea{\begin{eqnarray}}
\def\eea{\end{eqnarray}}
\def\ee{\end{equation}}

\begin{document}

\preprint{ FERMILAB-PUB-97/209-A, hep-ph/yymmnn}

\title{On the Nature of the Magnetic Fields Generated\\
During the Electroweak Phase Transition}
\author{Dario Grasso$^{(1)}$ and Antonio
Riotto$^{(2)}$}
 \address{$^{(1)}${\it Department of Theoretical Physics, Uppsala University,
 Box 803, S-751 08 Uppsala, Sweden}}
\address{$^{(2)}${\it NASA/Fermilab Astrophysics Center, \\ Fermilab
National Accelerator Laboratory, Batavia, Illinois~~60510-0500}}
\date{June 1997}
\maketitle
\begin{abstract}
In this Letter  we reanalyse  the question of  the origin   
 of magnetic fields during the electroweak phase transition. We show  that their formation  is  intimately connected  to    
some semiclassical configurations of the gauge
fields,  such as electroweak $Z$-strings and $W$-condensates. We describe the formation of  
these semiclassical configurations during a  first order phase  transition and argue that  they might be generated also  in the case of a second order phase transition. 
 We suggest  that the instability of electroweak strings does not 
imply the disappereance of the embedded magnetic field.
\end{abstract}

\baselineskip=20pt
\newpage
\section{Introduction}

An essential feature of phase transitions taking place in the early Universe is  the breaking of  translational invariance.
In first order phase transitions translational invariance is broken 
by   the nucleation of bubbles, while in second order phase transitions by the  formation of  
domains where  the order parameter is correlated.
Although some nontrivial  remnant of the breaking of translational invariance may 
 survive, analogously to the case of  a ferromagnet below the Curie
temperature, this is not generally the case in quantum  field theories where
 a uniform value of the order parameter is
energetically preferred. One remarkable  exception is represented by   
 topologically stable  defects  whose interiors do not feel the 
symmetry breaking and are formed by the Kibble mechanism \cite{Kibble}. 
Typical examples are 
local strings or domain walls.
However, the topology of the vacuum manifold in the electroweak model 
does not allow the presence of  topologically stable defects and one could wonder whether any trace of the structure present during the electroweak phase transition (EWPT) may remain imprinted  and eventually be  detectable
today. It has been suggested by Vachaspati  that the answer to this fundamental question might be positive \cite{Vacha91}. He suggested that  strong magnetic fields may be produced during
the electroweak phase transition  as a consequence of  nonvanishing spatial
gradients of the classical value of the Higgs field. These gradients arise due  to  the   finite correlation
length of the two-point correlation function just below the critical temperature. 
Once magnetic fields are generated they can  be imprinted in the highly conductive
medium and eventually survive. 
 Recently, Vachaspati's  suggestion has been  questioned \cite{Sacha}. The  electric current due the dynamics of the Higgs 
field has been computed and showed to be  vanishing during the EWPT.  It was concluded that  long range coherent magnetic fields are not generated by the
classical rolling of the   Higgs vacuum expectation value during the electroweak phase transition.  
In ref. \cite{Sacha}, however,  the contribution to the electric current
coming from the dynamics of the gauge fields was not considered since  it was  assumed that the classical value of these fields was vanishing. 
We think that such an assumption is not motivated. Indeed, we will show that  classical currents of the gauge fields,  and hence electromagnetic fields, are generally produced during the EWPT. 

It is the purpose of this Letter to reanalyse and possibly clarify  the question of  the origin   
of magnetic fields during the EWPT. 
We will argue that magnetic fields are indeed   formed during the EWPT and that their origin may be interpreted  to arise from the   appearance of 
some semiclassical configurations of the gauge
fields, such as electroweak $Z$-strings and $W$-condensates. 
The seeds of these configurations are the nonvanishing covariant derivatives of the Higgs field present during the phase transition.

We will  describe  the formation of  semiclassical gauge configurations 
during a first order  EWPT by bubble collisions.  This was already analysed by
Copeland and Saffin in \cite{SafCop2}, but here we extend their findings
with particular attention to the formation of the magnetic fields. In those cases
in which electroweak strings are formed
the equilibration of the  Higgs phases proceed in some analogy to   
 the $U(1)$ abelian toy model studied in \cite{KibVil,SafCop1}  
and magnetic fields may be  formed as a consequence of such  a process. 

By making use of some similarities of the electroweak model with
the superfluid $^3$He system, we will argue that electroweak strings are also expected to be formed if the EWPT is of the second order. 
Although   electroweak strings are unstable,  their decay does  not imply the disappereance of the 
embedded magnetic field. This effect may increase the chances for the magnetic
field to survive  thermal fluctuations.

\section{The importance of gauge field configurations.}

For sake of clarity, we briefly repeat  Vachaspati's argument for the generation of magnetic fields during the EWPT \cite{Vacha91}. The basic object to analyse is the  generalised 
 electromagnetic field tensor given by 
\begin{eqnarray}
\label{HooftSM}
F^{{\rm em}}_{\mu\nu} &\equiv& - \sin{\theta_{W}}{\hat \phi}^a(x)F^{a}_{\mu\nu} +
\cos{\theta_{W}}F^{Y}_{\mu\nu}\nonumber\\ 
&-&
i\frac {\sin{\theta_{W}}}{g}\frac{4}{\Phi^{\dag}\Phi} \left[
\left(D_{\mu}{\Phi}\right)^{\dag}D_{\nu}{\Phi} -  
D_{\mu}{\Phi}\left(D_{\nu}{\Phi}\right)^{\dag} \right], 
\end{eqnarray}
where 
\be
{\hat \phi}^{a} \equiv
\frac{\Phi^{\dag}\tau^{a}\Phi}{\Phi^{\dag}\Phi}~.
\ee 
The definition (\ref{HooftSM}) is inspired by the analogous t'Hooft 
 definition given for the Georgi-Glashow model \cite{t'Hooft}.
The remarkable feature  is that it is explicitly 
gauge invariant and reduces to the standard definition in the presence of a 
uniform Higgs background. Vachaspati observed that,  even if the gauge fields vanish,  
the second term in (\ref{HooftSM}) may remain nonzero due to the nonvanishing
gradient in the classical value of the Higgs field during the EWPT phase transition.
Of course, one can always make a gauge 
transformation to render  the gradients in the Higgs phase  vanishing.
However, this operation will induce nonvanishing   gauge fields and  expression    (\ref{HooftSM}) remains  unchanged.
Some ambiguity is already present at this level though. 
Indeed, as we transfer all the informations about the Higgs gradients into the gauge fields, 
it is not clear whether and eventually  how  the electromagnetic fields spring from the dynamics of the 
the gauge fields and, furthermore,  if the dynamics of the Higgs field can be completely decoupled. Some further inspection of the gauge field dynamics is certainly necessary  to answer these crucial questions. 
Using the equations of motions of the field strength tensors $F^a_{\mu\nu}$ and 
$F^Y_{\mu\nu}$, it is easy to show that 
\begin{eqnarray}
\label{MaxwellSM}
&&\partial^\mu F^{\rm{em}}_{\mu\nu} = -\sin\theta_W \left\{
D^\mu {\hat \phi}^a F^a_{\mu\nu} \right.\nonumber\\
&+&\left.
\frac{i}{g}\partial^\mu \left[\frac{4}{\Phi^{\dag}\Phi}\left(
\left(D_{\mu}{\Phi}\right)^{\dag}D_{\nu}{\Phi} -  
D_{\mu}{\Phi}\left(D_{\nu}{\Phi}\right)^{\dag} \right)\right]\right\}~, 
\end{eqnarray}
where $D_{\nu}{\hat \phi}^a = \partial_{\nu}{\hat \phi}^a +
g \epsilon^{abc} W^b_\nu{\hat \phi}^c$.
A useful exercise in order to clarify the physical nature
of the several contributions to the electric current sustaining the magnetic fields is to 
 imagine a region of space where  the electroweak symmetry  is broken everywhere. Because of gauge invariance, we can fix     the 
unitary gauge for the Higgs field. This  implies the reduction ${\hat \phi}^a = - \delta^{a3}$ 
and amounts to transfer all the physical informations from the Higgs field phases into the gauge fields. 
In this gauge Eq. (\ref{MaxwellSM}) reads
\begin{eqnarray}
\label{MaxwellUG}
\partial^{\mu} F^{\rm{em}}_{\mu\nu} &=&
+ie\left[ W^{\mu \dag}\left(D_\nu W_\mu\right) -
W^{\mu}\left(D_\nu W_\mu\right)^{\dag} \right]\nonumber\\
&-& ie\left[ W^{\mu \dag}\left(D_\mu W_{\nu}\right) -
W^{\mu}\left(D_\mu W_\nu\right)^{\dag} \right]  \nonumber \\ 
&-& ie \partial^\mu\left(W_\mu^{\dag} W_\nu - W_\mu W_\nu^{\dag} \right)\nonumber\\
&+& 2\tan\theta_W ~\partial^\mu\left(Z_\mu\partial_\nu\ln\rho(x) -
Z_\nu\partial_\mu\ln\rho(x)\right)~.
\end{eqnarray}
Here $\rho$ indicates the modulus of the Higgs field. 
The first  two  terms on the right-hand side of this equation 
are the $W$ convective terms and the third term is called the   spin term being   
 related to the $W$ anomalous magnetic moment \cite{AmbOle}. 
It is known  that these  terms can induce  an anti-screening 
of the external magnetic field \cite{AmbOle}. These terms
are of course classically vanishing in the absence of a $W$-condensate.
As we will show in more details below,
the last  term in (\ref{MaxwellUG}) is  also  related to some possible semiclassical
configurations for the  $Z$-field.  Hence,  as far as only the gauge sector of the electroweak theory is considered,  expression (\ref{MaxwellUG}) tells 
us that the currents sustaining classical electromagnetic fields have to reside in
nontrivial semiclassical configurations of the gauge fields. 

A gauge invariant electric current was computed in \cite{Sacha} and
showed to vanish. Clearly, such a  result is compatible with (\ref{MaxwellSM},\ref{MaxwellUG}) 
only in the case semiclassical configurations of the gauge fields are absent.
However, as we are going to show, this is generally not the case during
the EWPT.  

\section{  Bubble collisions in $SU(2)$.}

To warm up, let us  first focus on the case in which   the collisions involve   two different  bubbles 
carrying a different phase  in the pure $SU(2)$ case. 
In practice we are fixing $\theta_W = 0$ in the Weinberg and Salam model.
Here the situation is quite peculiar since no "electromagnetic" field is generated in spite of the presence of gradients in the Higgs field. 
The Higgs phase is assumed to be uniform across any single bubble.  
Following ref.\cite{SafCop2},  we may write the initial Higgs field configuration
as a superposition of the two independent bubbles separated by a space 
distance ${\bf b}$
\be
\label{phiin}
\Phi_{{\rm in}}({x}) = \frac{1}{\sqrt{2}}\left(
\begin{array}{c} 0\\ \rho({\bf x}) \end{array}\right) + 
\frac{1}{\sqrt{2}}\:\exp\left(-i \frac \theta 2 n^a\tau^a\right)
\left(
\begin{array}{c} 0\\ \rho({\bf x} - {\bf b}) \end{array}\right)~.
\ee
Eq. (\ref{phiin}) certainly provides a good description of the real physical
situation when the two domains are well separated and the mutual interaction may be neglected.
We assume that it  holds until the two bubbles collide and fix conventionally  $t = 0$ when the collision takes place. 
The  configuration  (\ref{phiin}) can be recasted  in the general form  $
\Phi_{{\rm in}}({\bf x}) = \frac{1}{\sqrt{2}} {\rm exp}\left(-i \frac {\tilde{ \theta}({\bf x})}{2} n^a\tau^a\right)
(0, \tilde{\rho})^T$ \cite{SafCop2}, 
where  the entire spatial dependence of the phase has been  factorized  
into a new phase $\tilde{\theta}({\bf x})$ (we will omit the tilde from now on). 
We assume that the gauge fields strength vanish before bubble collision. 
We also  impose that the initial gauge fields $W_\mu^a$  and their derivatives 
are zero at $t = 0$.
This condition is of course gauge dependent and should be interpreted as a gauge choice.

In vectorial form ${\hat  \phi}^a$  may be written as  $
{\hat  \phi} = \cos\theta~{\hat \phi}_0 +
\sin\theta~{\hat n}\times {\hat  \phi}_0 +
2 \sin^2{\frac \theta 2}~\left({{\hat n}}\cdot 
{\hat \phi}_0\right)~{{\hat n}}$, 
where $ {\hat  \phi}^T_0 \equiv -(0,0,1)$ . Note that we are now working 
in the adjoint representation for the Higgs field.
It is straightforward to verify that in the unitary gauge,  
 $\theta = 0$, ${\hat \phi}$ reduces to  ${\hat \phi}_0$.
Since the versor ${\hat n}$ associated to the $SU(2)$ gauge rotation does not 
depend on the space coordinates, we have the freedom to choose ${\hat n}$ to be everywhere perpendicular to 
${\hat {\phi}}_0$. In such a  case ${\hat {\bf \phi}}$ can be always obtained 
by rotating  the unit vector by an angle $\theta$ in the   
plane identified by  $\hat{n}$ and ${\hat {\bf \phi}}_0$. Formally, 
$
{\hat  \Phi} = \cos\theta {\hat \phi}_0 +
\sin\theta~{\hat n}\times {\hat \phi}_0$, 
which  clearly describes a simple  $U(1)$ transformation. This already suggests
  that such particular choice of the relative orientation of ${\hat n}$ and 
${\hat \phi}$ the dynamics of the  system is determined by an effective 
$U(1)$ and not by the entire $SU(2)$ gauge group. 
However, in order to verify this property more properly, we need to 
investigate the dynamics of the gauge fields. The latter 
is described by the equation of motion $
D^\mu F^a_{\mu\nu} = g |\rho|^2~\epsilon^{abc} D_\nu{\hat \phi}^b 
{\hat \phi}^c$ which, at   $t = 0$, reads  
\begin{equation}
\label{Weqin}
\partial^\mu F^{a}_{\mu\nu} = - g|\rho|^2 \partial_{\nu}\theta(x) 
\left(n^a - n^c {\hat \phi}^a{\hat \phi}^c\right)~.  
\end{equation}
If we now impose the  condition $\hat{n}\perp {\bf {\hat \phi}}_0$, it is straightforward to verify that $\hat{n}\perp {\bf {\hat \phi}}$. 
As a result, Eq. (\ref{Weqin}) reduces to 
\begin{equation}
\label{Weqin2}
\partial^\mu F^a_{\mu\nu} = - g|\rho|^2 \partial_{\nu}\theta(x) n^a~.  
\end{equation}
As anticipated,  only the gauge field component 
along the direction  $\hat{n}$,
namely $A_\mu=n^aW^a_\mu$, does posses some initial dynamics in virtue of the 
presence of a nonvanishing gradient of the phase between the two domain. In 
other words, the only  field strength possessing some dynamics is the one   
associated to the $U(1)$ gauge field $A_\mu$. 
The  interaction of critical bubbles during a first order phase transition in 
a pure $SU(2)$ theory  may be effectively described  by a simple  $U(1)$ gauge group.
As noted in  ref.\cite{SafCop2}, this case is of particular
interest as it may give rise to the formation of $W$ closed strings
during bubble collisions.   

To better address the issue of the formation of the "electromagnetic" fields,  we make use of t'Hooft definition of the electromagnetic field 
  for a pure $SU(2)$ gauge group  $
F^{{\rm em}}_{\mu\nu} \equiv - {\hat \phi}^aF^{a}_{\mu\nu} +
\frac {1}{g} {\hat \phi}^a
D_{\mu}{\hat \phi}^b D_{\nu}{\hat \phi}^c \epsilon^{abc}$ \cite{t'Hooft}. 
Since we are not considering the full electroweak
gauge group structure, it is understood here that $F^{{\rm em}}_{\mu\nu}$
is not the conventional electromagnetic field strength.
After some algebra one can verify that the condition ${\hat n}\perp {\hat \phi}$
implies $F^a_{\mu\nu}$ identically vanish (more technical details will be 
given in \cite{noi}). 
Hence, in the absence of stable topological defects such as monopoles, no electromagnetic fields
are produced during bubble collision even if the Higgs field has a nonvanishing gradients.  In other words, there are no currents to  sustain the "electromagnetic" field. 
This shows as the presence of nonvanishing gradients in the Higgs field is
not a sufficient condition for the generation of electromagnetic fields to 
take place.

\section{ Bubble collisions in the electroweak theory}

We now generalise the  previous discussion  
to  the gauge group  $SU(2)_L\otimes U(1)_Y$ of the
electroweak theory. We have to 
introduce an extra generator, the hypercharge with the relative phase $\varphi$. The generalisation of the form (\ref{phiin}) is straightforward.  
The gauge field equations  at $t = 0$ are given by
\begin{eqnarray}
\label{WeqSM0}
\partial^\mu F^a_{\mu\nu} &=& -\frac{g}{2} \rho^2(x) \left[- n^a 
\partial_\nu\theta +  {\hat \phi}^a \partial_\nu\varphi \right],\nonumber\\
\partial^\mu F^Y_{\mu\nu} &=& -\frac{g'}{2} \rho^2(x) \left[- n^a{\hat \phi}^a  
\partial_\nu\theta + 2 \partial_\nu\varphi \right].
\end{eqnarray}
Due to the presence of an extra generator with respect to the pure $SU(2)$ case, 
the reduction to a simple $U(1)$ is no  longer possible, but in some special  cases.   Different orientations
of the versor ${\hat n}$ with respect to ${\hat \phi}_0$ 
correspond to different physical situations, but in general both $W$- and $Z$-configurations are expected to form. 
Let us briefly address two extreme cases. If  ${\hat n}$ is orthogonal to 
${\hat \phi}_0$,   this implies
${\hat n}\perp {\hat \phi}$. As a consequence, at $t = 0$ 
we have on the right-hand side of Eq.  (\ref{WeqSM0}) the sum of two different 
terms that are perpendicular to each other. This means that
at least two independent generators will be involved
in the dynamics of the $SU(2)$ gauge fields. Thus, in general we
cannot reduce ourselves to an effective $U(1)$. Such a reduction would be
possible only imposing the additional assumption $\partial_\nu\varphi = 0$. 
In such a case the hypercharge field does not evolve and dynamics of the system reduces to that of a  pure $SU(2)$.  
Under such conditions $W$-strings may 
be formed \cite{SafCop2}. However, even if symmetry may be locally restored,
we have shown that no electromagnetic fields are produced in this case. 

The case   
in which $\hat{n}$ is parallel to  $\hat{\phi}_0$ is much more interesting. In such  a case 
${\hat \phi} = {\hat \phi}_0$ . We obtain 
\begin{eqnarray}
\partial^\mu F^3_{\mu\nu}&=&  \frac{g}{2} \rho^2(x)\left(\partial_\nu\theta+ \partial_\nu\varphi\right),\nonumber\\
\partial^\mu F^Y_{\mu\nu} &=& - \frac{g'}{2} \rho^2(x) \left(\partial_\nu\theta+ \partial_\nu\varphi\right).
\end{eqnarray}
No initial evolution for the $a = 1,2$ components of $F^a_{\mu\nu}$ is present. 
The equation of motion for the $Z$-field strength $F^{Z}_{\mu\nu} =  \cos{\theta_{W}}F^{3}_{\mu\nu}
 - \sin{\theta_{W}}F^{Y}_{\mu\nu}$ reads
\begin{equation}
\label{Zeq0}
\partial^\mu F^Z_{\mu\nu} =  \frac{\sqrt{g^2 + g^{'2}}}{2}\rho^2(x)
\left(\partial_\nu\theta+ \partial_\nu\varphi\right).
\end{equation}
This equation tells us that a gradient in the phases of the Higgs field gives rise to a nontrivial dynamics of the $Z$-field with an effective gauge coupling constant $\sqrt{g^2 + g^{'2}}$. Notice that this takes place even if $\partial_\mu\rho=0$. 
Thus, in agreement with \cite{SafCop2}, we have an effective 
reduction of the full $SU(2)\otimes U_Y(1)$ gauge structure to 
an abelian $U(1)$ group, at least at initial time.
The equilibration of the phase $(\theta+\varphi)$ can be now treated in          
analogy to the $U(1)$ toy model studied by Kibble and Vilenkin \cite{KibVil}, 
the role of the $U(1)$ "electromagnetic" field being now played by the
$Z$-field. Fixing an axial gauge for this field, with the $z$-axis chosen
along the line joining the bubble centres, it is easy to show that the 
only nonvanishing components of $F^{Z}_{\mu\nu}$ are a longitudinal
$Z$-electric field and a "ring-like" azimuthal $Z$-magnetic field. 
The related $Z$ field winds in planes normal to the ring internal axis.
An important difference with respect to \cite{KibVil} is that 
one does not  need to  require the radial part of the Higgs 
field to be spatially uniform and constant in time. Indeed, numerical 
simulations clearly indicate that $\rho$ has a nontrivial evolution during
bubble collisions \cite{SafCop1,SafCop2}. This is crucial not only for
for the violation of the geodesic rules, 
but also for magnetic field generation. 
Let us take for simplicity $\partial_\mu\theta=0$. The complete set of equations of motion we may   write at finite,
though small, times is
\begin{eqnarray}
\label{nieole}
&&\partial^\mu F^{Z}_{\mu\nu} =  \frac{g}{2\cos\theta_W}\rho^2(x)
\left(\partial_\nu\varphi + \frac{g}{2\cos\theta_W} Z_\nu\right),\nonumber\\
&&d^\mu d_\mu\left(\rho(x) {\rm e}^{i\frac{\varphi}{2}}\right) + 
2\lambda \left(\rho^2(x) - \frac{1}{2}\eta^2\right)\rho(x){\rm e}^{i\frac{\varphi}{2}}=0,
\end{eqnarray}
 where $d_\mu = \partial_\mu + i\frac{g}{2\cos\theta_W} Z_\mu$, $\eta$ is the vacuum expectation value of $\Phi$ and $\lambda$ is the quartic coupling.
Note that, in analogy with \cite{KibVil}, a  gauge invariant phase difference can be introduced 
by making use of the covariant derivative $d_\mu$.
Equations (\ref{nieole}) are the Nielsen-Olesen equations of
motion  \cite{NieOle,Vacha93}. Their solution describes 
a $Z$-vortex where $\rho=0$ at its core.
The  reader should keep  in mind that, as
follows from our previous considerations, the geometry of the  problem
implies that the vortex is closed, forming a ring which axis coincide
with the conjunction of bubble centres. What is crucial is that the formation of the magnetic field is always associated to the appearance of a semiclassical $Z$-configuration. Indeed, $\partial^\mu F^{\rm em}_{\mu\nu}$ does not vanish:  even rotating away the phase $\varphi$  
\begin{equation}
\partial^\mu F^{\rm em}_{\mu\nu}= 2\tan\theta_W \partial^\mu\left(
\tilde{Z}_\mu\partial_\nu \ln{\rho(x)} -  \tilde{Z}_\nu\partial_\mu \ln{\rho(x)} \right)
\end{equation}
 where now $\tilde{Z}_\mu$ is $Z$-field in the new gauge. What is important is that $\tilde{Z}_\mu$ has a nontrivial dynamics. 
A ring-like magnetic field is formed along the 
internal axis of the vortex. It is interesting to observe that if closed 
$Z$-vortices break into finite segments, {\it e.g.} due to thermal
fluctuations or subsequent bubble collisions, a magnetic flux will emanate
from the segment's extremities which will behave as a pair of magnetic 
monopoles. This effect was already suggested in \cite{Vacha94}. 

Magnetic fields were ignored in \cite{SafCop2}. However, their role
is crucial for the late evolution of the $Z$-vortices and the 
surviving of the $U(1)$ reduction. In fact,
as the magnetic field induce a back-reaction on the charged gauge fields,
it is clear that the formation of the magnetic field in the core of the 
$Z$-string spoil the  reduction of the $SU(2)\otimes U(1)_Y$ group to 
an effective $U(1)$. Together with the restoration  
of the electroweak symmetry in the core of the string, the magnetic field 
induces the decay of the $Z$ string into a $W$-condensate \cite{Perkins}.
While electroweak symmetry restoration in the core of the string reduces  $m_W$, the magnetic field via its coupling 
to the anomalous magnetic moment of the $W$-field, causes, for $eB > m_W^2$, 
the formation of a condensate of the $W$-fields.
The presence of a $W$-condensate gives rise to an electric 
current which  can sustain magnetic 
fields even after the $Z$ string has disappeared. This may have relevant 
consequences on the subsequent evolution of magnetic fields and we leave this 
investigation for future work \cite{noi}. 

It is important to notice that, in the most general case,  $\hat{n}$ is neither parallel nor perpendicular to $\hat{\phi}_0$ and we expect the formation of nontrivial $W$- and $Z$-configurations \cite{noi}. In such a case, one should retain the non-abelian nature of the electroweak theory and no  reduction to a simple $U(1)$ abelian group is expected to hold. 

We can now wonder what is the strength of the magnetic fields at the end 
of the EWPT. A partial answer to this question has been recently given in ref.     \cite{AhoEnq} where the formation of  ring-like magnetic fields in collisions of
 bubbles of broken phase in an abelian Higgs model were inspected. 
Under the assumption that magnetic fields
are generated by a process that resembles the Kibble and Vilenkin 
\cite{KibVil} mechanism, it was concluded that  a  magnetic field of the order of  $B  \simeq 2 \times 10^{20}$ G with a coherence length  
of about $10^2~\rm{GeV}^{-1}$ may be originated. Assuming turbulent enhancement of the field by inverse
cascade, 
a root-mean-square value of the magnetic field $B_{{\rm rms}} \simeq 10^{-21}$ G   on a comoving scale of $10$ Mpc should be present today \cite{AhoEnq}. 
Although our previous considerations give some partial support to the scenario 
advocated in \cite{AhoEnq} we have to stress, however, that only in some restricted  cases it is possible to reduce the dynamics of the system to the dynamics of a simple $U(1)$ abelian group. Furthermore, once $Z$-vortices are formed
the non-abelian nature of the electroweak theory shows due to the back-reaction
of the magnetic field on the charged gauge bosons
and it is not evident  that the same numerical values  obtained in \cite{AhoEnq} will be obtained in the case of the EWPT.  This and other issues, {\it e.g.} how likely 
is it to form loops, what distribution should we expect and on what length scale, will be addressed in a separate publication \cite{noi}.

\section{Magnetic fields from a second order transition.}

Let us now briefly address  the formation of electromagnetic fields in the case  case in which 
the EWPT is second order. 
As we argued, electromagnetic fields which  are not 
 merely  thermal fluctuations can only be formed in the presence of 
semiclassical gauge field configurations. If the EWPT transition is of the  second order,  domains where the Higgs field is 
physically correlated appear near  the critical temperature.
Although these correlated domains have  properties quite different from the  
bubbles formed during a first order transition, it is however plausible that 
gauge field configurations can  be formed  during a second order transition too. 
The formation of vortices is a
common phenomenon in second order phase transitions.  In particular, 
$^3$He to $^3$He-A and $^3$He to $^3$He-B second order phase transitions
are known to give rise to the formation of topological and non-topological 
vortices via the  Kibble mechanism. It is known that non-topological vortices 
in these systems share many common aspects with the electroweak 
strings \cite{VolVac}.  The use of condensed matter physics experiments to 
investigate the non-perturbative aspects of particle physics and the formation
of defects in the early-Universe is a very modern and active research line
(see \cite{Zurek} for a review). We adopt the same point of view to argue 
that electroweak strings are actually formed during the EWPT if this is second 
order.  
 
In order to estimate the density of vortices, hence the mean magnetic 
field,  we need to determine the typical size of
domains. A very reasonable estimate of the typical minimum size
of the domains in the vicinity of the critical temperature is given by the
correlation length of the Higgs field computed at the temperature at which
thermal equilibrium between the false-vacuum (which is now the symmetric
phase $\phi=0$) and the true-vacuum is no longer attained. In other words,
we are interested in the temperature at which thermal fluctuations of the
Higgs field inside a given domain of broken symmetry are no longer able to
restore the symmetry. This is basically the Ginzburg criterion to
determine what is generally called the Ginzburg temperature $T_G$. A very
{\it rough} estimate of $T_G$ may be obtained just equating the thermal
energy $\sim T$ with the energy contained in a domain of size $\ell$ of
broken phase, $E_\ell\simeq \lambda \:\langle\phi(T)\rangle^4\:\ell^3$,
where $\ell$ is typically taken to be the correlation length $\xi(T)$ and $\lambda$ is the quartic coupling in the Higgs potential.
Here, however, we need a more precise determination of the Ginzburg
temperature and, in this respect, we will follow the criterion suggested
in ref. \cite{KolGle}. Let us imagine that a domain of broken
symmetry has been formed in the vicinity of the critical temperature and
that the value of the Higgs field inside is of order of $\langle
\phi(T)\rangle$. We may model a thermal fluctuation which restores the
symmetry inside the domain ({\it i.e.} the symmetry is unbroken, or
$\langle \phi(T)\rangle=0$, in a sub-region of the domain) by a
sub-critical bubble having the following configuration 
\begin{equation}
\phi_{{\rm ub}}(r)=\langle \phi(T)\rangle\left(1-{\rm
e}^{-r^2/\ell^2_{{\rm ub}}}\right), 
\end{equation} 
where $\ell_{{\rm ub}}$ is the correlation length in the symmetric phase. 
The rate per unit volume and unit time of nucleating such a sub-critical 
bubble of symmetric phase inside a domain of broken phase (with size 
equal to the correlation length in the broken phase) may be estimated to be 
\begin{equation}
\Gamma_{{\rm ub}}=\frac{1}{\ell_{{\rm b}}^4}\:{\rm e}^{-S_3^{{\rm ub}}/T},
\end{equation}
where $\ell_{{\rm b}}$ is the correlation length in the broken phase.  
$S_3^{{\rm ub}}$ is the high temperature limit of the Euclidean action 
computed in correspondence of the configuration given in Eq. (13) and 
is a complicated function of the parameters present in the Higgs effective 
potential. A complete expression for $S_3^{{\rm ub}}$ may be found in 
ref. \cite{Enqvist} and we do not give it here. We would like only to  
notice  that, at fixed 
$T$,  $S_3^{{\rm ub}}/T$ increases as $\lambda$ increases (and the phase 
transition becomes very weak first order or second order), rendering the
thermal fluctuations less and less efficient as it might be  conjectured 
by making use  of the Ginzburg criterion outlined above. 

Thermal fluctuations of the 
unbroken phase inside a domain of broken phase freeze out and cease to
be nucleated when the rate $\Gamma_{{\rm ub}}$ becomes smaller than
$H^4$, $H$ being the Hubble expansion rate of the Universe. This happens 
when $S_3^{{\rm ub}}/T\simeq {\rm ln}(M_{P\ell}^4/T_c^4)\simeq 160$.
We have numerically computed the temperature $T_G$ at which thermal 
fluctuations freeze out for different values of the parameter $\lambda$ 
(or equivalently for different values of the physical Higgs boson mass 
$M_H$) and checked that nucleation of regions of unbroken phase
inside a domain of broken phase stops at temperatures very close to the 
critical temperature, $T_G=T_c$ within a few percents. The corresponding
size of the domain of broken phase is determined by the correlation length
in the broken phase at $T_G$
\begin{equation}  
\frac{1}{\ell(T_G)^2_{{\rm 
b}}}=V^{\prime\prime}\left(\langle\phi(T_G)\rangle,T_G\right)
\end{equation}
and is weakly dependent on $M_H$, $\ell_{{\rm b}}(T_G)\simeq 11/T_G$ for
$M_H=100$ GeV and $\ell_{{\rm b}}(T_G)\simeq 10/T_G$ for $M_H=200$ GeV. 
 
Using this result and eq. (\ref{HooftSM}), we may  estimate 
a magnetic field of order of $B \sim 4e^{-1} \sin^2\theta_W  \ell_{{\rm b}}^2(T_G)
\sim 10^{22} ~\rm{G}$, 
 on a correlation length $\ell_{{\rm b}}(T_G)$.  Notice that this value is about two 
orders of magnitude smaller than the one suggested by Vachaspati in his original paper \cite{Vacha91}, the reason being that the correlation scale adopted there was about one 
order of magnitude larger the one obtained here by detailed balance arguments.  
The computation of the root-mean-square value of the magnetic field on scales larger
that  $\ell_{{\rm b}}(T_G)$ would require an estimate of the probability of formation of the  
$Z$-strings. This and  other issues like the stability,  the strength and the spatial distribution of the magnetic fields at the end of the EWPT 
 are currently under investigation \cite{noi}.

We conclude that it is plausible that magnetic fields are produced
during the EWPT as a consequence of a nontrivial dynamics of the gauge fields.
As classical magnetic fields are odd both under $C$ and $CP$, 
it is noticeable that this process give rise to spontaneous breaking of both
these symmetries.   
   
\vskip 1cm
\centerline{\bf Acknowledgements}

The authors would like to thank E.~Copeland, A.~Dolgov, G.~Ferretti, P.~Olesen,
H.~Rubinstein and O.~T\"{o}rnqvist for several valuable discussions.
D.G. is grateful  to the N.~Bohr Institute, Copenhagen,
and to the CERN Theory Division for hospitality. The work of D.G. was partially supported
by the NorFA grant No.97.15.049-O.  
A.R. is supported by the DOE and NASA under grant NAG5--2788.

\def\apj#1#2#3{{\it Astrophys.\ J.\ }{{\bf #1} {(#2)} {#3}}}
\def\app#1#2#3{{\it Astropart.\ Phys.\ }{{\bf #1} {(#2)} {#3}}}
\def\np#1#2#3{{\it  Nucl.\ Phys.\ }{{\bf #1} {(#2)} {#3}}}
\def\pr#1#2#3{{\it Phys.\ Rev.\ }{{\bf #1} {(#2)} {#3}}}
\def\pl#1#2#3{{\it  Phys.\ Lett.\ }{{\bf #1} {(#2)} {#3}}}
\def\prl#1#2#3{{\it Phys.\ Rev.\ Lett.\ }{{\bf #1} {(#2)} {#3}}}
\def\prep#1#2#3{{\it Phys.\ Rep.\ }{{\bf #1} {(#2)} {#3}}}
\def\jp#1#2#3{{\it J.\ Phys.\ A\ }{{\bf #1} {(#2)} {#3}}}


\begin{references}

\bibitem{Kibble}
        T.W.B.~Kibble, \jp {9}{1976}{1387}.

\bibitem{Vacha91}
        T.~Vachaspati, \pl {B265}{1991}{258}.

\bibitem{Sacha} S.~Davidson, \pl {B380}{1996}{253}.

\bibitem{SafCop2} P.M.~Saffin and E.J.~Copeland, hep-th/9702034.

\bibitem{KibVil}
        T.W.B.~Kibble and A.~Vilenkin,
        \pr {D52}{1995}{679}.


\bibitem{SafCop1} P.M.~Saffin and E.J.~Copeland, 
        \pr {D54}{1996}{6088}.

\bibitem{t'Hooft}
        G.~t'Hooft, \np {B79}{1974}{276}.

\bibitem{AmbOle} J.~Ambjorn and P.~Olesen,
        {\em Int. J. Mod. Phys.}{\bf A5}, (1990) 4525.

\bibitem{noi} E.J.~Copeland, D.~Grasso, A.~Riotto and P.M.~Saffin, in preparation. 

\bibitem{NieOle} H.B.~Nielsen and P.~Olesen,
        \np {B61}{1973}{45}. 

\bibitem{Vacha93} T.~Vachaspati, 
        \prl {68}{1992}{1977}.

\bibitem{Vacha94}  T.~Vachaspati, ``Electroweak Strings,
Sphalerons and Magnetic Fields'' in the proceedings of
``Electroweak Physics and the Early Universe'', eds. J. C. Romao 
and F. Friere, Plenum Press, New York, 1994; hep-ph/9405286.

\bibitem{Perkins} W.B.~Perkins, 
        \pr {D47}{1993}{R5224}.


\bibitem{AhoEnq} J.~Ahonen and K.~Enqvist, hep-ph/9704334.

\bibitem{VolVac} G.E.~Volovik and T.~Vachaspati, cond-mat/9510065.

\bibitem{Zurek} W.H.~Zurek, 
        \prep {276}{1996}{177}.

\bibitem{KolGle} E.W.~Kolb and M.~Gleiser, 
        \np {B364}{1991}{411}.

\bibitem{Enqvist} K.~Enqvist et {\it  al}, 
        \pr {D45} {1991} {3415}.

\end{references}
\end{document}